\documentclass[rnote]{aa}
\usepackage{graphicx}
\usepackage{txfonts}

\begin{document}

\title{On the multiwavelength spectrum of the microquasar 1E~1740.7$-$2942}
\authorrunning{Bosch-Ramon et al.}
\titlerunning{On the spectrum of 1E~1740.7$-$2942}

\author{V. Bosch-Ramon\inst{1} \and 
G. E. Romero\inst{2,3,} 
\and J.~M. Paredes\inst{1} \and A. Bazzano\inst{4} \and M. Del Santo\inst{4} 
\and L. Bassani\inst{5}}

\institute{Departament d'Astronomia i Meteorologia, Universitat de Barcelona, Av. 
Diagonal 647, 08028 Barcelona, Catalonia, Spain; vbosch@am.ub.es, jmparedes@ub.edu.
\and Instituto Argentino de Radioastronom\'{\i}a, C.C.5,
(1894) Villa Elisa, Buenos Aires, Argentina; romero@iar.unlp.edu.ar.
\and Facultad de Ciencias Astron\'omicas y Geof\'{\i}sicas, UNLP, 
Paseo del Bosque, 1900 La Plata, Argentina.
\and INAF - Istituto di Astrofisica Spaziale e Fisica cosmica di Roma,
via del Fosso del Cavaliere 100, 00133 Roma, Italy
\and INAF - Istituto di Astrofisica Spaziale e Fisica cosmica di Bologna, via
Gobetti 101, 40129 Bologna, Italy}


\abstract
{The microquasar 1E~1740.7$-$2942 is a source located in the direction of the Galactic
Center. It has been detected at X-rays, soft gamma-rays, and in the radio band, showing an extended
radio component in the form of a double-sided jet. Although no optical counterpart has been found so far
for 1E~1740.7$-$2942, its X-ray activity strongly points to a galactic nature.}{We aim to improve our
understanding of the hard X-ray and gamma-ray production in the system, exploring whether the jet 
can emit significantly at high energies under the light of the present knowledge.}{We
have modeled the source emission, from radio to gamma-rays, with a cold-matter dominated jet model. 
INTEGRAL data combined with radio and RXTE data, as well as EGRET and
HESS upper-limits, are used to compare the computed and the observed spectra.}{From our modeling, 
we find out that jet
emission cannot explain the high fluxes observed at hard X-rays without violating at the same
time the constraints from the radio data, favoring the corona origin of the hard X-rays. 
Also, 1E~1740.7$-$2942 might be detected by GLAST or
AGILE at GeV energies, and by HESS and HESS-II beyond 100 GeV, with the spectral 
shape likely affected by photon-photon absorption in the disk and corona photon fields.}{}

\maketitle

\section{Introduction} \label{intro}

1E~1740.7$-$2942, discovered at X-rays by the {\it Einstein} satellite (Hertz \& Grindlay 1984)  and located at less
than one degree from the Galactic Center, is considered to be the first microquasar since  the observation of radio
jets extending $\sim 1'$ (Mirabel et~al. 1992). There are hints of correlation between the radio and the strong X-ray
emission (Mirabel et al. 1993), and X-ray state changes have been observed with timescales similar to those of other
X-ray binaries (Del Santo et~al. 2005 and references therein). Moreover, a 12.7-days periodical modulation of a 3--4\%
in the X-ray emission seems to be produced by the orbital motion of the system  (Smith et al. 2002). This X-ray
behavior suggests that the source is galactic in nature. The modulation period and the upper-limits in magnitude
obtained in the infrared band  suggest that the stellar companion in 1E~1740.7$-$2942 might be a low-mass star
(Mart\'i et~al. 2000). The behavior of the X-ray spectrum, very similar to those of other galactic black-hole
candidates (or BHC) (e.g. Sunyaev et~al. 1991), points to a black-hole as the compact object. The inferred star mass,
the strong X-ray radiation and the little X-ray orbital modulation, hinting perhaps to a circular orbit, point to
accretion via Roche-lobe overflow. The fact that the companion has not been detected yet is likely due to the dense
surrounding medium, with high concentrations of dust and a large hydrogen column density of about $10^{23}$~cm$^{-2}$
(Gallo \& Fender 2002), which depletes photons from the infrared up to soft X-ray energies. The same authors also 
detected hints of an extended X-ray halo. This  hydrogen column  density is even higher than the typical value in the
direction of the Galactic Center ($\sim 10^{22}$~cm$^{-2}$; Dickey \& Lockman 1990), and might imply that this object
is embedded within a molecular cloud (Yan \& Dalgarno 1997). This peculiar location  could also explain the very
extended structures detected at different wavelengths around this source, as was pointed out by Mirabel et~al. (1992).
Regarding the distance to 1E~1740.7$-$2942, we have adopted in this work the Galactic Center distance, 8~kpc.

Concerning the emission properties, the source shows spectrally flat and faint radio emission coming from the core
with flux densities of less than 1~mJy (Fender 2001 and references therein;  $\sim 10^{29}$~erg~s$^{-1}$ at 8~kpc), as
well as optically thin radio emission coming from the radio lobes. In the 10-100 keV band, {\it RXTE} and  {\it
INTEGRAL} observed the source in August 2003 when it was in the low-hard state finding fluxes of about
$10^{-9}$~erg~s$^{-1}$~cm$^{-2}$ ($\sim 10^{37}$~erg~s$^{-1}$; Del Santo et~al. 2005). Further data have been taken by
{\it INTEGRAL},  which altogether with the previous ones have been reduced and analyzed for this work with the most
recent software package (OSA~5.1), obtaining very similar results to those already published. EGRET observed
intensively the Galactic Center region in the range $\sim 0.1-10$~GeV. No source was detected at the position of 
1E~1740.7$-$2942, which provides upper-limits for its emission at these energies. At the Galactic Center distance the
inferred luminosity upper-limits are about $10^{35}$~erg~s$^{-1}$ (Hartman et al. 1999). Moreover, the flux of the
nearby source 3EG~J1744$-$3011 can be taken as an absolute upper-limit\footnote{The EGRET sensitivity in that region
is confuse, as it is seen in Fig.~3 of Hartman et~al. (1999), and the angular resolution is poor all along the
Galactic plane.}, being about $10^{-10}$~erg~s$^{-1}$~cm$^{-2}$ ($\sim 10^{36}$~erg~s$^{-1}$). The Cherenkov telescope
HESS, working above  0.1~TeV has not detected 1E~1740.7$-$2942, although significant diffuse emission has been found
in the surroundings of the Galactic Center (Aharonian et al. 2006) that could cover intrinsic TeV emission from  this
source. It is probably premature to state that this source cannot be detected by HESS at very high energies with longer
exposures, or by HESS-II in the near future. 

We want to explain the emission of this object in the context of microquasars during the low hard state 
(Fender et~al. 2003), when emission at radio and higher energies from a compact jet is produced and
accretion processes are likely contributing or even dominant at X-rays. Data in the radio band and X-rays/soft
gamma-rays (from {\it RXTE} and {\it INTEGRAL})\footnote{{\it
CHANDRA} data from the observations performed by Gallo \& Fender (2002) is not  considered here due to
its non-simultaneity with the observations carried out by {\it RXTE} and {\it INTEGRAL}. We note
however that at that time the flux at few keV was about one order of magnitude lower than the flux
presented here in  similar energy ranges. We note that radio is either non-simultaneous, but only moderate 
radio flux variations have been found (Mirabel et~al. 1993).}, and upper-limits at high-energy and very high-energy
gamma-rays (EGRET and HESS), have been considered. We have applied a detailed jet model to
explain the data and to make predictions at high energies, and to give some constraints about
the physics of the processes involved in producing the broadband emission.

\section{A jet model applied to 1E~1740.7$-$2942}

A model based on a freely expanding magnetized jet, in which internal energy is dominated by a cold proton plasma
extracted from the accretion disk, has been developed by Bosch-Ramon et~al. (2006) for microquasars in the low-hard
state. The energetics of accretion, ejection, and radiation is treated consistently. We fix the jet Lorentz factor to be
1.25, since microquasar jets during the low-hard state seem to be mildly relativistic (see, e.g., Gallo et al. 2003). We
adopt a jet viewing angle of 45$^{\circ}$. To keep the jet dominated by cold matter, the magnetic field energy density
is taken below the equipartition value. Internal shocks occur within the jet accelerating a fraction of the jet leptons
up to very high energies with a power-law energy distribution. The density of relativistic leptons is  kept low to avoid
their internal energy dominance, fixing the relativistic lepton to cold proton number maximum ratio to 1/10.
Relativistic protons are not treated here, although some amount could be present in the jet, constrained by the cold
matter dominance condition (concerning their radiative properties, see e.g. Romero et~al. 2005). The acceleration rate
is taken $\le 0.1\times q_{\rm e}Bc$ (see Protheroe 1999), which is balanced by electron energy losses 
and the accelerator size determining the
maximum electron energy. The minimum energy of the non-thermal electrons 
has been determined equaling their gyroradius with that of protons of temperature $\sim 10^{11}$~K, 
which dominate in our model the jet internal energy. In
the context of a low mass system, as 1E~1740.7$-$2942 seems to be, the accelerated leptons radiate via synchrotron
emission and, at higher energies, via inverse Compton (IC) scattering with synchrotron (SSC) and disk/corona
photons. The disk and corona photon densities decrease inversely
with the square of the distance when seen from the jet. However, photons coming from the disk X-ray emitting region, pretty close to the compact object,
are considered as coming from behind the jet, whereas photons from the corona
are treated at this stage as isotropic\footnote{The corona and the jet base, where the corona IC is more efficient, 
could overlap.}. A multicolor
blackbody spectrum is adopted for the disk, and a power-law plus an exponential cutoff for the corona (see Del Santo
et~al. 2005). Photon-photon absorption in the dominant photon fields is computed. We
have also taken into account roughly the IC emission from the created pairs to compute the spectral energy
distributions (SED). IC scattering and gamma-ray absorption in the low-mass star photon field
are negligible.
All the parameter values are summarized in Table~\ref{tab}. Those related to the disk and the corona radiation are taken
from Del Santo et~al. 2005, and those regarding the jet geometry are fixed to typical values found in the literature
(e.g. Bosch-Ramon et~al. 2006).
We have assumed a black-hole
mass of 5~M$_{\odot}$.

\section[]{Results}\label{res}

We have explored three parameters: the fraction of the accreted matter that is ejected in form of (two)
jets ($q_{\rm jet/accr}$), the acceleration efficiency, and the shock energy dissipation efficiency. For
this parameter study, the electron power-law index is fixed to a typical value in Fermi I acceleration
theory, 2.2. In general, a high $q_{\rm jet/accr}$ implies low dissipation rates since otherwise it
would overcome observed radio fluxes, reducing the chances of TeV detection but rendering still a good
GeV emitter. A smaller $q_{\rm jet/accr}$ allows for larger dissipation efficiencies, thus increasing
the very high-energy emission. Too small $q_{\rm jet/accr}$ does not yield radio fluxes at the observed
level, let alone detectable gamma-ray fluxes. The explored parameter space and related features are
summarized in Table~\ref{tab} (bottom). 

\begin{figure}
\resizebox{\hsize}{!}{\includegraphics{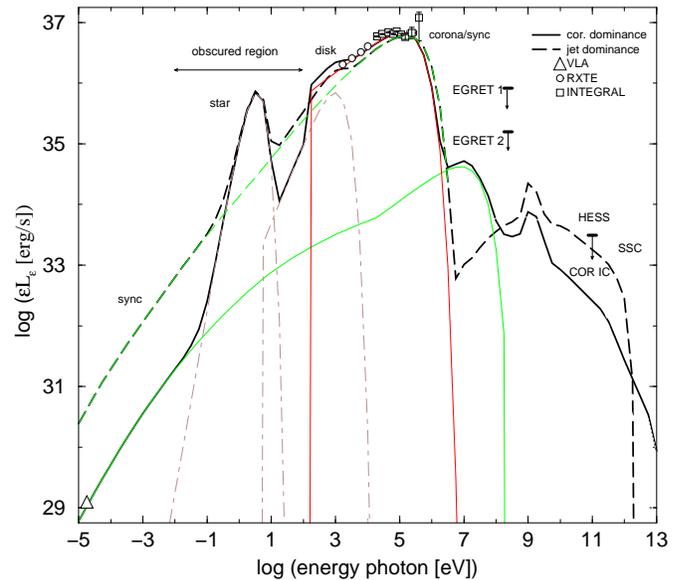}}
\caption{
Computed broad-band SED for 1E~1740.7$-$2942 for a dominant X-ray corona (thick solid line) and a dominant X-ray jet 
(thick long-dashed line) (see Table~\ref{tab2}). Data points and observational upper-limits are also shown. Note 
that radio constraints are violated for the dominant X-ray jet case.
We note that dips at $\sim 100$~MeV and few GeV are due to photon-photon absorption
in the corona and the disk photon fields respectively. At the radio
band, the core luminosity at the adopted distance is shown. The slope at this range is similar to that observed 
(not shown explicitly; see Fender 2001 and references therein). At X-rays/soft  gamma-rays, {\it RXTE} and {\it
INTEGRAL} data are plotted (Del Santo et~al. 2005). At high-energy gamma-rays, two upper-limits are shown: one from 
the spectrum of the source 3EG~J1744$-$3011 (EGRET 1), as absolute upper-limit  (Hartman et~al. 1999), and the other
obtained from the sensitivity limits of EGRET (EGRET 2) in the region (Hartman et~al. 1999). The diffuse emission flux
at $\ge$100~GeV observed  by HESS in the direction of the source is shown as an upper-limit (Aharonian et~al. 2006).}
\label{plot1}
\end{figure}

Next, we explore two representative cases: the corona dominates at X-ray energies (Del Santo et~al. 2005); X-ray
emission comes mainly from the jet (Markoff et~al. 2001; Georganopoulos et~al. 2002). The parameter values adopted in
each individual case are shown in Table~\ref{tab2}. In Fig.~\ref{plot1}, we show computed SEDs for both cases. 
When the
corona dominates, the jet is adopted to be mild, with high dissipation and acceleration efficiencies (to explore 
whether gamma-ray emission is significant). We have neglected here
those components that are not relevant to model the SED of this particular source. The accretion rate,
1.2$\times10^{-8}$~M$_{\odot}$~yr$^{-1}$, has been fixed assuming that the total X-ray luminosity (corona+disk) is
0.05$\times c^2$ the total accretion rate. The corresponding Eddington accretion rate for a 5~M$_{\odot}$ black hole
is 2$\times 10^{-7}$~M$_{\odot}$~yr$^{-1}$.  The magnetic field at the base of the jet\footnote{The magnetic field
decreases like $1/z$  as far as cold matter energy density goes down along the jet.} and the (one) jet total kinetic
luminosity are about $4\times 10^4$~G and $4\times10^{36}$~erg~s$^{-1}$, respectively. The expected fluxes are several
10$^{-13}$~erg~cm$^{-2}$ s$^{-1}$ at 1~GeV, and close to 10$^{-13}$~erg~cm$^{-2}$ s$^{-1}$ at 100~GeV. The source
would have not been detected above 100~MeV, since it is below the lower EGRET limit, and its contribution to the
nearby source could not be significant. HESS upper limit is at the moment above the computed emission level. 
 
For a dominant synchrotron X-ray jet (see Fig.~\ref{plot1}), we adopt more extreme parameter values than those explored above and neglect the
corona to
try to reproduce the broadband emission. The
synchrotron modeling of the X-rays requires a two-sided jet with $q_{\rm jet/accr}$=0.2, energy dissipation efficiency
of 70\%, acceleration efficiency of 0.002 and an accretion rate taken to be a 10\% of the Eddington accretion rate
quoted above  (i.e. 2$\times10^{-8}$~M$_{\odot}$~yr$^{-1}$). The magnetic field and jet total kinetic luminosity are
$3\times 10^5$~G and $3\times10^{37}$~erg~s$^{-1}$. Particularly for this case, the magnetic field has been increased
to reach the observed X-ray fluxes. For the dissipation efficiency in this case, the pure radiative efficiency is
very large, of about a 50\%. Radio fluxes are exceeded by more than one order of magnitude, and an electron power-law
index of 1.7 is required. Instead of being the result of synchrotron emission, the X-rays might be produced via IC
scattering. When attempting to reproduce the observed X-ray spectrum with an IC jet model with a weak corona, it is
not possible to reach the X-ray fluxes through IC of external photons and/or SSC emission because the jet power
requirements are too high, and it is not possible to keep the  particle energy low enough as to make them radiate just
up to a few hundreds of keV for any reasonable acceleration rate with any photon field. In addition, for the SSC
model, radio constraints are also violated. We conclude, in the context of our model and after exploring a vast
range of parameter values, that the corona X-ray dominated emission reproduces better the observed broadband SED than
the jet X-ray dominated emission in 1E~1740.7$-$2942.

Finally, we have explored semi-quantitatively the radiation from the extended radio emitting jets. 
A magnetic field of 10$^{-4}$~G all along that jet region and a jet carrying at least 1\%
of the accreted matter was required in our model to generate 
radio emission up to 2-3 pc, i.e. the size of the radio lobes detected by Mirabel et~al.
(1992) at 8~kpc. This magnetic field is similar to what is typically found in molecular
clouds (Crutcher 1999) and above the equipartition value with jet matter in those regions. We recall that
1E~1740.7$-$2942 could be located within a molecular cloud (Yan \& Dalgarno 1997).

\begin{table}[]
  \caption[]{Parameter values}
  \label{tab}
  \begin{center}
  \begin{tabular}{ll}
  \hline
   Parameter [units] & value \\
 \hline
Disk inner part temperature [keV] & 0.5 \\
Disk luminosity [erg~s$^{-1}$] & $1.5\times10^{36}$ \\
Corona photon index & 1.6 \\
Corona emission peak [keV] & 300 \\
Disk inner radius [R$_{\rm Sch}$] & 50 \\
Jet initial point (compact object RF) [R$_{\rm Sch}$] & 50 \\
Jet semi-opening angle tangent & 0.1 \\
Max. ratio hot to cold lepton number & 0.1 \\
Compact object mass [M$_{\odot}$] & 5 \\
Jet Lorentz factor & 1.25 \\
Jet viewing angle [$^{\circ}$] & 45 \\  
  \end{tabular}  
  \begin{tabular}{llll}
  \hline
$q_{\rm jet/accr}$ & accel. eff. & diss. eff. & prediction \\
 \hline
0.2 & 0.1 & 0.01 & up to TeV \\
0.2 & 10$^{-5}$ & 0.003 & up to GeV \\
0.04 & 0.1 & 0.1 & "bright" at 100~GeV\\
0.01 & 10$^{-5}$ & 0.1 & up to 100~MeV\\
  \hline
  \end{tabular}
  \end{center}
\end{table}

\begin{table}[]
  \caption[]{Specific parameter values for different models}
  \label{tab2}
  \begin{center}
  \begin{tabular}{lll}
  \hline
   Parameter [units] & corona  & jet \\
 \hline
Accretion rate [M$_{\odot}$/yr] & $1.2\times10^{-8}$ & $2\times10^{-8}$ \\
Corona luminosity [erg~s$^{-1}$] & $3.5\times10^{37}$ & $10^{35}$ \\
$q_{\rm jet/accr}$ & 0.04 & 0.2 \\
Shock dissipation efficiency & 0.1 & 0.7 \\
Acceleration efficiency & 0.1 & 0.002 \\
Electron power-law index & 2.2 & 1.7 \\
Equipartition parameter & 0.1 & 0.3 \\
  \hline
  \end{tabular}
  \end{center}
\end{table}

\section{Discussion and summary}

There is a wide debate in the astrophysical community about the possible origin of the X-rays in
microquasars  since, at the present state of knowledge, the corona and the jet scenarios seem to be
roughly consistent in some cases with observations (Markoff et~al. 2005). In 1E~1740.7$-$2942, the hard
X-ray emission is difficult to be explained if coming from a jet since it would imply an energetic
efficiency in the jets significantly larger than for the corona emission (see Sect~\ref{res}). Actually,
in the context of our model, synchrotron emission from the jet seems to be able to explain the low-hard
state X-ray spectrum of the source, but it exceeds largely the observed core radio fluxes.
1E~1740.7$-$2942 appears to show related radio and X-ray variability (Mirabel et al. 1993), although is
radio underluminous in the context of the radio-X-ray luminosity correlation found for black hole
candidate X-ray binaries and associated with the accretion/ejection activity (Gallo et~al. 2003, Corbel
et~al. 2003, Fender et~al. 2003). This might be due to a particularly radiatively efficient corona. We
remark that radio variations (Mirabel et~al. 1993) could hardly explain a flux as large as that
predicted with the synchrotron jet model  (there remains the possibility that other  jet models, with a
much higher electron minimum energy, may be consistent with observations). All this points to the fact
that, although accretion and jet phenomena are probably linked, the dominant component at X-rays can be
either the jet or the disk/corona depending on the source.

After the discovery of gamma-ray emission from microquasars (Paredes et~al. 2000, Aharonian et al. 2005,
Albert et al. 2006), gamma-rays are becoming of great interest as a probe for X-ray binaries. In this
context, we remark that jets in general, if sharing similar properties, are likely to emit above X-rays,
reaching in some cases very high energies. Our model predicts significant amounts of gamma-rays from
1E~1740.7$-$2942 that might be detected by future space- and ground-based instruments like {\it GLAST},
{\it AGILE} or HESS and HESS-II, although photon-photon absorption in the disk and corona fields could
prevent these detections, as can be seen comparing  the two cases explored here (see Fig.~\ref{plot1};
for more precise calculations showing the importance of the discussed effects, see e.g. Akharonian \&
Vardanian 1985, Wu et~al. 1993, Bednarek 1993). 

Concerning the large-scale jets, Gallo \& Fender (2002) stated that the magnetic field should be $\ge
1$~$\mu$G, in agreement with our semi-quantitative  results but inferred on the basis of completely
different grounds. Our estimate of the magnetic field would imply that no significant IC gamma-rays can
be originated in that region, since the synchrotron emission channel would be more efficient than the IC
one. It would not prevent high-energy emission to be produced via other mechanisms, like hadronic
radiative mechanisms possibly at work within the mentioned molecular cloud  that could surround
1E~1740.7$-$2942 (see, e.g., Bosch-Ramon et al. 2005).

\begin{acknowledgements}
J.M.P. and V.B-R. acknowledge partial support by DGI of the
Ministerio de Educaci\'on y Ciencia  (Spain) under grant AYA-2004-07171-C02-01, as well as additional
support from the European Regional Development Fund (ERDF/FEDER). During this work, V.B-R has been
supported by the DGI of the Ministerio de (Spain) under the fellowship BES-2002-2699. G.E.R is
supported by the Argentine Agencies CONICET (PIP 5375) and ANPCyT (PICT 03-13291).
A.B., L.B and MDS acknowledge partial support by ASI under contract
I/R/046/04.
\end{acknowledgements}

{}

\end{document}